\documentclass[
 amsmath,amssymb,
 aps, prl, twocolumn, superscriptaddress,citeautoscript
]{revtex4-2}

\usepackage{mathtools}
\usepackage{amsmath,amssymb}
\usepackage{graphics,graphicx,color,calc}
\usepackage{xcolor}
\usepackage{dcolumn}
\usepackage[english]{babel}
\usepackage{mathrsfs}
\usepackage[mathcal]{eucal}
\usepackage{bm}
\usepackage{lipsum}
\usepackage{xr}
\usepackage{stackengine}
\usepackage{soul}

\usepackage{xcolor}
\newlength\raisedepth
\def\[{\left[}
\def\]{\right]}
\def\({\left(}
\def\){\right)}

\usepackage{bookmark}

\newcommand{\UGA}
{\affiliation{Universit\'e Grenoble Alpes, CNRS, LIPhy, 38000 Grenoble, France}}
\newcommand{\rri}
{\affiliation{Soft Condensed Matter Group, Raman Research Institute, Bangalore 560080, Karnataka, India}}

\begin{document}


\title{{\it{Trainable amorphous matter}}: tuning yielding to rigidity via mechanical annealing}



\title{{\it{Trainable amorphous matter}}: tuning yielding by mechanical annealing}

\author{Maitri Mandal}
 \thanks{These authors contributed equally to this work.}
 \rri
\author{Pappu Acharya}
 \thanks{These authors contributed equally to this work.}
 \UGA
\author{Rituparno Mandal}
\email{rituparno@rri.res.in}
\rri

\author{Sayantan Majumdar}
\email{smajumdar@rri.res.in}
\rri

\begin{abstract}

Living organisms can demonstrate highly adaptable and sophisticated responses using ‘memory’ resulting from repeated exposure to external conditions or ‘training’. However, realizing similar adaptability in mechanical responses in inanimate, physical materials presents an outstanding challenge in several fields, including soft matter, materials science, and in the domain of soft robotics, to name a few. Our study focuses on disordered solids, which are model systems that resemble granular matter, foam and other disordered, soft solids. Here, combining bulk rheology, in-situ optical imaging, and numerical simulations, we demonstrate how training 
via cyclic shear can encode memories that tunes the yield point in a unique way and over unprecedented ranges. 
Our study reveals that such tunability is intricately linked to the plasticity, non-affine deformations, and formation of shear bands. Remarkably, our numerical simulations illustrates that systems with identical internal energies, prepared via different protocols (mechanical or thermal), can display markedly different  rheological responses, indicating that energy alone does not determine mechanical behaviour. Moreover, while the yield strain increases with training amplitude, the material simultaneously softens, contrasting with the thermal case where both quantities increase monotonically with increasing annealing. Our results open up possibilities for memory-induced tuning of mechanical response in {\it{trainable amorphous matter}}, independently or in combination with thermal annealing, far beyond the material--feature space achievable via the latter alone.

\end{abstract}

\maketitle

\section{Introduction}
 
 Disordered solids are found everywhere in our daily lives and have important scientific and industrial applications~\cite{DeGennes_RevModPhys_1999,Berthier_RevModPhys_2011,Bonn_RevModPhys_2017,Zaiser_NatRevPhy_2023,Moon_JAP_2025}. Examples range from soft materials such as pastes, gels, foams, and emulsions, to structural materials like silica glass, concrete, asphalt, and compressed soils. Despite their diverse origins and compositions, these systems share a common character: their behavior, be it dynamical or mechanical, depends strongly on how they were produced and how they have been treated subsequently~\cite{Kob_PRL_1997,Sastry_Nature_1998, Fiocco_PRL_2014,Misaki_PNAS_2018,Rev.Mod.Phys_2019,Yeh_PRL_2020,Barlow_PRL_2020,Edera_PRX_2025}. This inherent `preparation history' dependence often makes their attributes challenging to understand, predict and control~\cite{Misaki_PNAS_2018, Yeh_PRL_2020,Bhaumik_PNAS_2021,Edera_PRX_2025} for desired applications.

A central feature of their mechanical response is the existence of a yield point, at which the material transitions from an elastic, solid-like state, that stores energy in elastic deformations, to a flowing, liquid like one~\cite{Maloney_PRE_2006,Leishangthem_NatCom_2017,Misaki_PNAS_2018,Lulli_PRX_2018,Berthier_NatRevPhy_2025}, where stored energy is dissipated through plastic rearrangements of its constituents. 
Therefore a yield point is a critical threshold that defines the maximum load that a material can sustain before it begins to deform irreversibly. In many contexts, a change in this threshold can have severe consequences. For instance, soils that appear stable under everyday conditions can, beyond certain stresses, lose their ability to support load and collapse catastrophically~\cite{Holzer_1999,Puzrin_2004}; thus shifting the yield point upward in such cases can significantly improve resistance to mechanical failure. Conversely, lowering the yield point can be desirable in processes where ease of flow and reshaping are required~\cite{Saeki_1998,Schall_Science_2007,Amani_2012}. Other salient aspects of the mechanical response of such disordered solids include material elasticity, characterized by different elastic moduli and the degree of brittleness.

\begin{figure*}[t]

 \includegraphics[height=9.8cm]{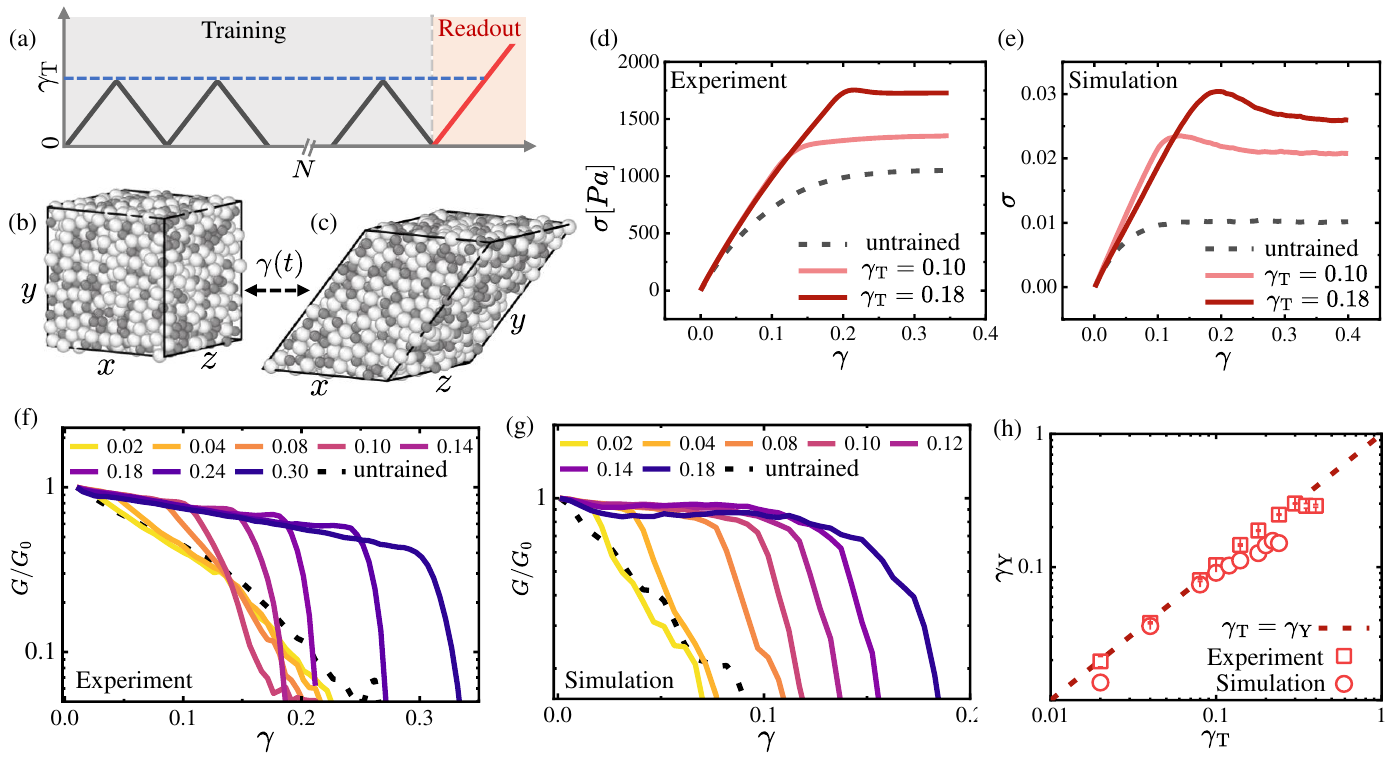}
 \centering
 \caption{\textbf{Training induced tunability of the yield point in model glasses.} (a) Schematic of the shear strain perturbation protocol applied to an amorphous solid during training and read-out: black lines represent the training cycles where strain is varied in an asymmetric cycle: $0 \rightarrow \gamma_{\mathrm{T}}\rightarrow 0$. After $N$ training cycles, a uniform strain is applied (shown in red) for read-out, which exceeds $\gamma_\mathrm{T}$. 
 The corresponding 3D representations of the un-sheared and sheared configurations from our particle based simulations are shown in (b) and (c) respectively. Stress-strain response of untrained and trained systems during readout in experiments (d) and simulations (e). Here we have used x-y component of the stress tensor and subtracted out any pre-stress in the system.
 In both cases, untrained systems show ductile yielding, while trained systems show more brittle-like yielding, occurring just beyond the training amplitude $\gamma_\mathrm{T}$. 
 This tunability of the yield point can be observed more clearly in the normalized differential shear modulus ($G/G_0$; see text for the definition) during readout in experiments (f) and simulations (g). 
 (h) The yield strain $\gamma_\mathrm{Y}$ (see the SI section S1 for the method of identifying the yield point) shows a one-to-one correlation with the applied $\gamma_\mathrm{T}$ over a wide range, both in experiments and simulations. For the experimental data the mean and error bars are computed from three independent samples.
 }
 \label{fgr:Yield_point_shift}
\end{figure*}
While changes in the shear modulus modify the stiffness landscape that a disordered system explores before plasticity sets in~\cite{Maloney_PRL_2004,Tsamados_PRE_2009,Hentschel_PRE_2011}, variations in brittleness determine whether failure occurs abruptly through rapid fracture or gradually via distributed plastic rearrangements~\cite{Xi_PRL_2005,Misaki_PNAS_2018,Bhaumik_PNAS_2021}.
Understanding and controlling these properties are therefore of paramount importance for designing materials that meet specific performance requirements ~\cite{Jang_NatMat_2010,Misaki_PNAS_2018,Divoux_SoftMatter_2024,Mutneja_2025_PRL}. However, designing materials with a prescribed chemical composition that simultaneously realize targeted yield strength, bulk modulus, and brittleness remains a fundamental challenge ~\cite{Jang_NatMat_2010,Logon_Materials_2021,Reichenauer_Langmuir_2022,Dasari_PNAS_2023}. The difficulty is especially pronounced in bulk amorphous solids, where the vast configurational space and intrinsic structural disorder impede direct control over mechanical response.


On the other hand, as indicated earlier, disordered solids are also known to exhibit a rich family of memory effects~\cite{Rev.Mod.Phys_2019,Nagel_JCP_2023,Paulsen_AnnRev_2025}, in particular in the context of mechanical training~\cite{Keim_PRL_2011,Fiocco_PRL_2014,Adhikari_EuroPhy_2018,Mukherji_PRL_2019,keim_PRR_2020,keim_SciAdv_2022,Lindeman_SciAdv_2025,Mungan_PRL_2025}. For example, when cyclically sheared at a fixed strain amplitude, a  disordered solid can evolve toward a reversible steady state in which particle rearrangements become minimal. During a subsequent readout, the mean-squared displacement  shows a pronounced dip at the training amplitude, demonstrating the ability of the material to remember previously applied deformation amplitude. 
One question therefore naturally arises, can we control the response of a  disordered material by encoding mechanical memory, in spirit of Wolf's law~\cite{Wolff_1986}? In other words can we create mechanical metamaterials, where different arrangement of particles (for an identical composition), facilitated via mechanical training, can have desired rheological response?~\cite{Stern_AnnRev_2023,Jaeger_SoftMatter_2024}

In this work, we show through experiments on colloidal glasses and particle based molecular dynamics simulations, that cyclic shear or mechanical annealing (MA), offers a precise and unique way to control the yield point over an unprecedented range of strain amplitudes. While conventional thermal annealing (TA) typically drives a material toward progressively more rigid and brittle state, MA produces a qualitatively different outcome: with increasing shear amplitude during training, the yield point tends to shift towards a predetermined strain, while the material softens and the degree of brittleness changes in a non-monotonic way. In addition, in situ imaging and particle-based simulations reveal that this training-induced shift in the yield point is accompanied by a pronounced increase in microscopic non-affine displacements once the system is strained beyond the training amplitude, indicating that the bulk response, such as brittle yielding, is closely tied to catastrophic collective particle rearrangements manifested via shear bands. 

Our findings further demonstrate that such variations in mechanical response are not solely dictated by the changes in the material’s internal energy. By contrasting MA and TA, we show that states with identical internal energy, when accessed through distinct annealing protocols, can exhibit strikingly different yielding behavior. Moreover, the combined use of thermal and mechanical annealing offer versatile routes to the  material design space that is fundamentally unattainable through thermal treatment alone. These synergistic mixed-annealing strategies thus 
opens up new possibilities for engineering functional materials. 

\section{Results}

\subsection{Tuning yield point via cyclic shear perturbation}

In our experiments, we study a model colloidal amorphous solid composed of colloidal (PNIPAM: poly(N-isopropylacrylamide)) particles dispersed in an aqueous medium. We perform rheological measurements in which both the training and the readout involve controlled shear strain perturbations. Detailed descriptions of the sample preparation, experimental setup, and perturbation protocol are provided in the Methods section. To complement our experiments we carry out simulations of 
 a bi-disperse dense assembly of soft harmonic spheres~\cite{Durian_PRL_1995,Durian_PRE_1997,Hern_PRL_2002,Hern_PRE_2003,Hecke_JOP_2009,Acharya_PRL_2020} and subject them to either athermal quasi-static shear~\cite{Maloney_PRE_2006,Karmakar_PRE_2010, Leishangthem_NatCom_2017} or thermal annealing~\cite{Misaki_PNAS_2018}, depending on the required protocol.
Full details of both simulation protocols are provided in the Methods section.

Fig.~\ref{fgr:Yield_point_shift}(a) shows a schematic of the memory encoding protocol deployed via asymmetric oscillatory shear perturbation, where triangular strain pulses ($0 \rightarrow \gamma_{\mathrm{T}}\rightarrow 0$) are applied over $N$ cycles both in experiments and in  particle based simulations. To investigate whether memory has been successfully encoded or not, we apply a uniform shear strain as a readout protocol. A typical unstrained and a strained configuration (from 3d particle based simulations) are shown in Fig.~\ref{fgr:Yield_point_shift}(b) and Fig.~\ref{fgr:Yield_point_shift}(c), respectively.

We observe that the rheological response of the trained system during the readout differs markedly from that of the untrained sample. While the untrained system fails to exhibit a pronounced linear regime in the stress–strain curve and shows a ductile or smooth yielding from the beginning, the trained sample displays a strengthened linear response, which becomes longer with increasing training amplitude, as shown in Fig.~\ref{fgr:Yield_point_shift}(d). This strengthening arises from particle rearrangements driven by a process akin to random organization ~\cite{Pine_Nature_2005,Corte_NatPhys_2008,Lundberg_PRE_2008,Royer_PNAS_2015,Lavrentovich_PRE_2017}, wherein repeated cyclic deformation leads to progressively lower configurational potential energy—effectively functioning as a form of mechanical annealing (MA). The yielding behavior of the trained sample also shows sharper (or brittle) nature, which is strikingly different from the untrained one. Remarkably, our simulations based on a minimal model of amorphous solid, show a very similar stress strain response (see Fig.~\ref{fgr:Yield_point_shift}(e) for comparison) as our experimental findings. 

\begin{figure*}[t!]

 \includegraphics[height=8.5cm]{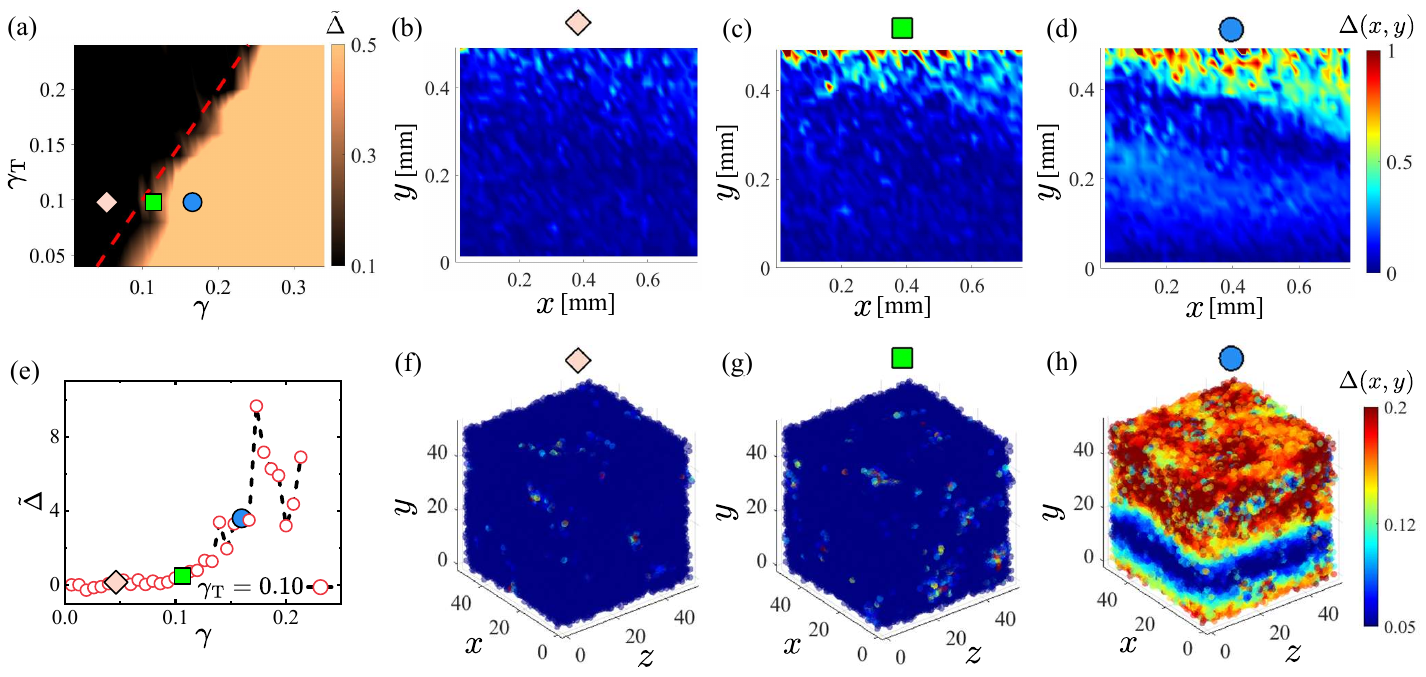}
 \centering
 \caption{\textbf{Emergence of non-affine motion during readout and shear banding.} (a) Relative non-affinity $\tilde{\Delta}$
 is plotted as a color map during the readout phase for various training amplitudes $\gamma_{\mathrm{T}}$. 
 For all values of $\gamma_{\mathrm{T}}$ the increase in relative  non-affinity begins predominantly when the applied strain $\gamma$ goes beyond the training amplitude $\gamma_{\mathrm{T}}$ (the red dashed line indicates $\gamma=\gamma_{\mathrm{T}}$). (b)--(d) Experimental non-affinity maps for a system trained at $\gamma_{\mathrm{T}} = 0.10$ are shown at different points along the readout:(b) at $\gamma= 0.05$ (below $\gamma_{\mathrm{T}}$, marked by a diamond), (c) at $\gamma= 0.11$ (just beyond $\gamma_{\mathrm{T}}$, marked by a square), and (d) at $\gamma= 0.16$ (well beyond $\gamma_{\mathrm{T}}$, marked by a circle). These maps visually capture the growth in non-affinity when the system is strained beyond the training amplitude. (e) Relative non-affinity $\tilde{\Delta}$ during readout from simulations for a system trained at $\gamma_{\mathrm{T}}= 0.10$, showing a similar onset of increase beyond the training amplitude. (f)--(h) Non-affinity maps from particle based simulations, corresponding to the strain values $\gamma = 0.05, \gamma=0.11,$ and $\gamma= 0.16$,
 provide a scenario similar to the one observed experimentally.
}
 \label{fgr:Nonaffinity}
\end{figure*}


The impact of the training is more pronounced in the differential modulus, computed from the stress strain response, measured during the readout. As shown in Fig.~\ref{fgr:Yield_point_shift}(f), systems trained within a broad range of amplitudes exhibit a sharp drop in the normalized differential modulus (where the differential modulus $G=d\sigma/d\gamma$ is normalized by its initial value $G_0$) right at the training strain $\gamma_{\mathrm{T}}$. This indicates that the system has retained a mechanical memory of the training amplitude, and yields precisely when that same strain is reached during readout. The mechanical response overall reflects the emergence of a well-defined elastic regime, bounded by a trainable and reproducible yield point. For the systems trained at amplitudes beyond this suitable window of training, the sharpness of the differential modulus response begins to degrade. The drop in $G/G_0$ becomes less rapid, signaling a loss of correspondence between training and yielding. Instead of exhibiting a well-defined threshold, the system in this case yields more gradually, progressively loosing the memory of the training amplitude. The corresponding response of the modulus $G$ in simulations, during uniform shear readout are shown in Fig.~\ref{fgr:Yield_point_shift}(g). Our particle based simulations reproduce key features observed in the experiments, including the training-dependent shift in the yield point and the sharp drop in the differential modulus.

The yield points $\gamma_{\mathrm{Y}}$, identified from the $G/G_0$ vs. $\gamma$ data (shown in Fig.~\ref{fgr:Yield_point_shift}(f) and Fig.~\ref{fgr:Yield_point_shift}(g)), are plotted against the applied shear strain or training amplitudes $\gamma_{\mathrm{T}}$ in Fig.~\ref{fgr:Yield_point_shift}(h). The method used to determine $\gamma_{\mathrm{Y}}$ is detailed in the  SI section S1. Notably, the yield point closely follows the applied training amplitude over a broad range—up to $\gamma_{\mathrm{T}}\approx0.3$ in experiments and $\gamma_{\mathrm{T}}\approx0.2$ in simulations—demonstrating a robust and tunable link between the training via oscillatory shear and the mechanical yielding. This direct correlation enables precise control over the yielding point across nearly one order of magnitude in strain amplitude, a level of tunability that is unprecedented to the best of our knowledge. Importantly, such a correlation between $\gamma_{\mathrm{T}}$ and  $\gamma_{\mathrm{Y}}$ does not depend on the specific form of cyclic shear used; the phenomenon persists even under a symmetric shear protocol $(0 \rightarrow \gamma_{\mathrm{T}}\rightarrow 0 \rightarrow -\gamma_{\mathrm{T}}\rightarrow 0)$ (see the SI section S2). 

\subsection{Emergence of enhanced non-affinity in cyclic shear beyond respective training amplitudes:}

As shown in the previous section, cyclic shear strain perturbations allow the yield point of the material to be tuned, with yielding occurring just beyond the respective training amplitude $\gamma_{\mathrm{T}}$. To examine whether this bulk response correlates with underlying microscopic dynamics, we perform in situ boundary imaging during the readout phase. Captured images are analyzed using {\it{Particle Image Velocimetry}} (PIV) to extract the experimental velocity field. By subtracting the corresponding affine part, we compute the non-affine displacement field $\Delta(x,y)$, where $x$ and $y$ correspond to the flow and velocity gradient directions, respectively (details are provided in the Methods section).

To quantify the dependence of non-affinity on training amplitude and readout strain, we define the relative measure $\tilde{\Delta}=(\overline{\Delta}/\overline{\Delta}_0 - 1)$, where $\overline{\Delta}$ is the spatial average of $\Delta(x,y)$ over the entire field of view in the $x$-$y$ plane, and $\overline{\Delta}_0$ is the mean baseline non-affinity at the beginning of the readout.
Fig.~\ref{fgr:Nonaffinity}(a) shows that $\tilde{\Delta}$ increases significantly beyond $\gamma_\mathrm{T}$ over the whole training range, indicating that macroscopic yielding is accompanied by the onset of enhanced non-affine dynamics (the evolution of the unnormalized mean non-affinity $\overline{\Delta}$ is provided in the SI section S3). This correlation suggests that the training-induced shift in the yield point is not only a bulk feature, but reflects underlying changes in the local/microscopic rearrangements that emerge once the system is strained beyond the training amplitude.

Fig.~\ref{fgr:Nonaffinity}(b)-- \ref{fgr:Nonaffinity}(d) show the spatial maps of non-affinity during readout for an experimental trained system with $\gamma_\mathrm{T}=0.10$. The maps correspond to: (b) $\gamma=0.05$ (diamond; below $\gamma_{\mathrm{T}}$), (c) $\gamma=0.11$ (square; around $\gamma_{\mathrm{T}}$), (d) $\gamma=0.16$ (circle; well beyond $\gamma_{\mathrm{T}}$). These maps illustrate the growth of non-affinity in the displacement as the system is strained progressively until the training amplitude. Non-affinity maps for other values of $\gamma_{\mathrm{T}}$ are provided in the SI section S4. We observe that such pronounced non-affinity can lead to the emergence of shear bands. These bands indicate regions where plastic rearrangements accumulate, producing large deformation and flow, that coexist with comparatively undeformed surrounding regions. Such localization represents a distinct mode of post-yield deformation, in which the bulk plastic response is carried predominantly by narrow zones of intense particle rearrangement rather than being distributed uniformly throughout the material~\cite{Langer_PRE_2001,Shi_PRL_2007,Manning_PRE_2009,Fall_PRL_2010,Dasgupta_PRL_2012,Fielding_RepProgPhy_2014,Parisi_PNAS_2017,Priezjev_Metals_2020,Pollard_PRR_2022}. Notably, the onset of shear bands is not simultaneous with the initial macroscopic yield but tends to occur at slightly higher strains during the post-yield regime. 
Similar observations have been made in simulations, where Fig.~\ref{fgr:Nonaffinity}(e) shows the evolution of a similar non-affine measurement ${\tilde{\Delta}}$ during readout for $\gamma_{\mathrm{T}}=0.10$, clearly illustrating the rapid increase in non-affine displacement following yielding. This quantity ${\tilde{\Delta}}$ is the average value of absolute displacement per particle (after subtracting the affine displacement from the shear deformation).
Fig.~\ref{fgr:Nonaffinity}(f) to \ref{fgr:Nonaffinity}(h) present the corresponding non-affine displacement maps for the simulations, at the same strain values as in the experiments, clearly highlighting the increase in non-affinity beyond $\gamma_{\mathrm{T}}$. More importantly, these maps also reveal the formation of shear bands, demonstrating that localized deformation emerges in both simulations and experiments after the system yields.

\begin{figure*}[t!]

 \includegraphics[height=10.5cm]{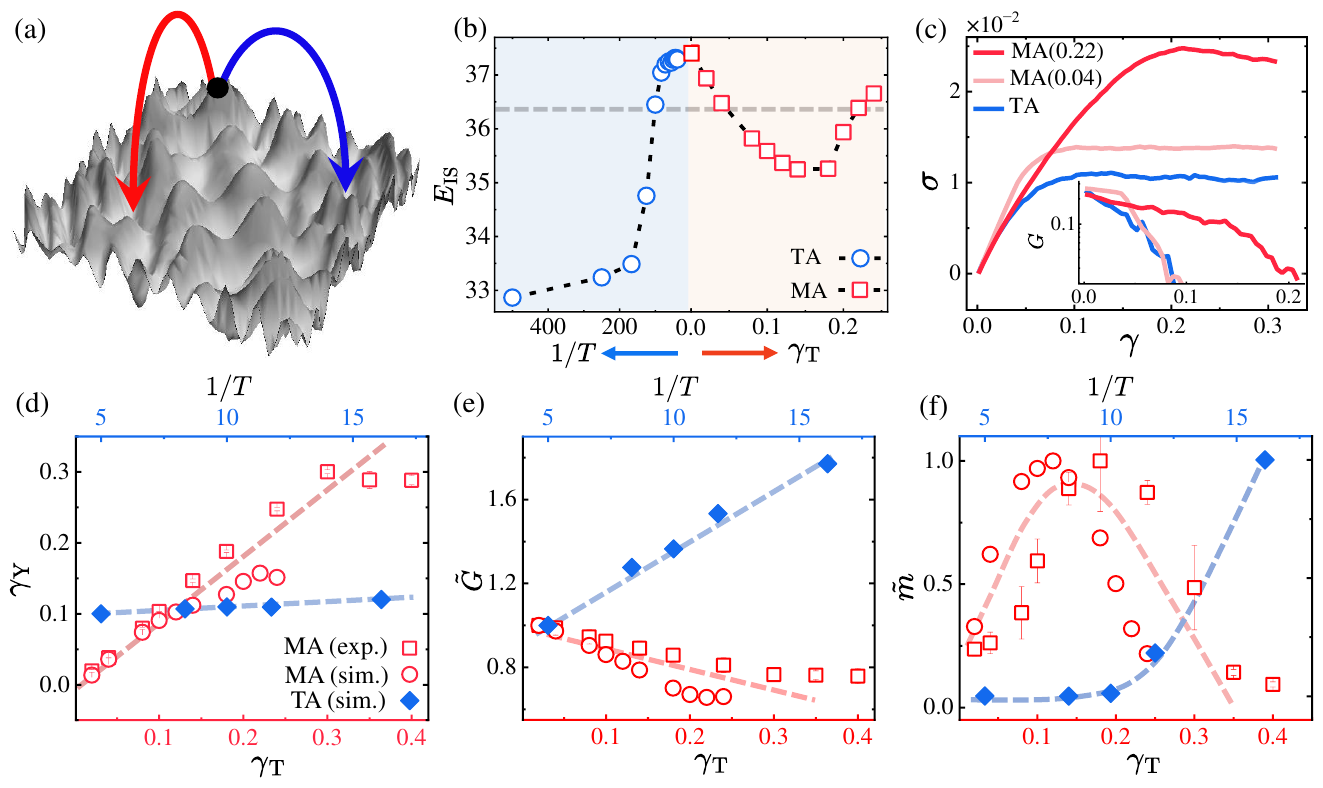}
 \centering
 \caption{\textbf{Comparison of thermal and mechanical annealing.} (a) Schematic of a rugged energy landscape, where the arrows represent annealing protocols achieved via mechanical and thermal means.
 (b) Total potential energy of the inherent structures ($E_\mathrm{IS}$) of thermally annealed systems as a function of inverse temperature (left, blue shaded region) and mechanically annealed systems as a function of shear amplitude (right, red shaded region); here both $T$ and $\gamma_\mathrm{T}$  controls the degree of annealing in respective cases. For TA, $E_\mathrm{IS}$ changes monotonically whereas, for MA we see a non-monotonic evolution of $E_\mathrm{IS}$. (c) Readout response (blue, light red and dark red represents the thermally annealed system,  mechanically annealed system with $\gamma_{\mathrm{T}}= 0.04$ and $\gamma_{\mathrm{T}}= 0.22$, respectively) under uniform shear for the three equi-energy (marked by a horizontal dashed grey dashed line in (b) sub-figure) configurations. Inset shows differential modulus corresponding to the stress–strain curves, highlighting distinct yielding behavior among these systems.
 For (d)--(f) blue diamond symbols represent TA data (from the Ref.~\cite{Misaki_PNAS_2018} extracted using PlotDigitizer software), red squares represent data from our experiments, and circles correspond to data from our MD simulations for MA. (d) Yield strain  $\gamma_{\mathrm{Y}}$,
 shows for TA only a mild increase in $\gamma_{\mathrm{Y}}$ while MA offers a broad tunability of the yield point across a wide range of $\gamma_{\mathrm{T}}$.
(e) Normalized shear modulus $\Tilde{G}$ (see the SI section S1 for details) shows an increasing trend with thermal annealing whereas mechanical annealing seem to make the material softer.
(f)  Normalized brittleness (see the SI section S1 for details)
exhibits peak brittleness at intermediate training amplitudes for MA, whereas TA shows a monotonic increase in brittleness. In all panels, red and blue dashed lines are guides to the eye for mechanical and thermal annealing trends, respectively. 
}
 \label{fgr:TA_MA_energy}
\end{figure*}

\begin{figure*}[t!]
\includegraphics[height=12cm]{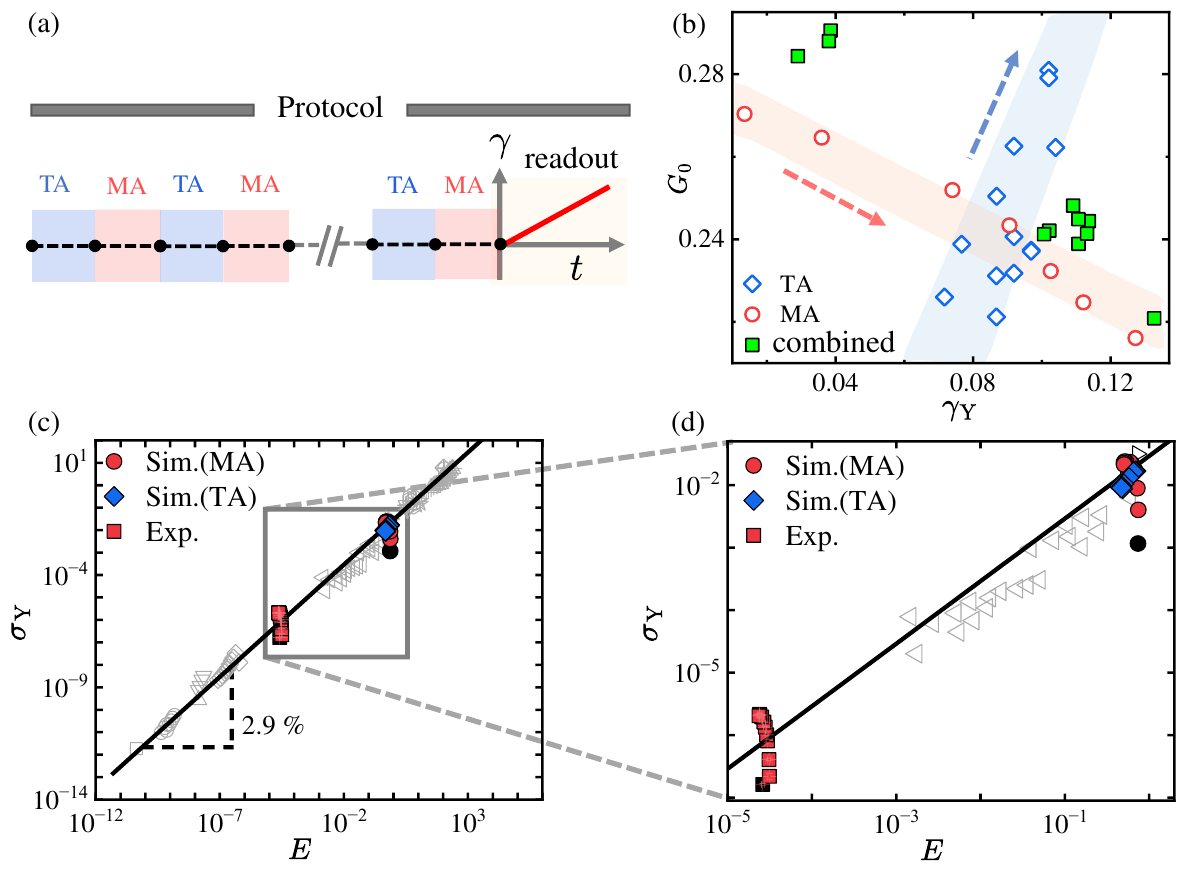}
 \centering
 \caption{\textbf{Material property tuning through thermo-mechanical annealing. }(a) Protocol for thermo-mechanical annealing, where sequences of thermal annealing (TA) and mechanical annealing (MA) 
 were applied before a uniform shear readout. (b) Initial shear modulus $G_0$ from simulations for TA (blue diamonds) and MA (red circles) shows stark difference in trends. Combined TA–MA sequences (green squares) reach states inaccessible to either TA or MA alone. (c) Yield stress $\sigma_{\mathrm{Y}}$ versus Young’s modulus $E$ for various disordered materials are digitized from the Ref.~\cite{Cubuk_science_2017} (open symbols), and the slope of the black line indicates the proposed universal yield strain 2.9\%. Our experimental results (solid red squares) and simulation results (solid red circles for MA and solid blue diamonds for TA) are overlaid. (d) Zoomed in view of (c), highlighting that mechanically annealed data points deviate from the dashed line along the vertical direction, indicating that mechanical annealing can defy universal mechanical response of known disordered materials. Data for the untrained systems are shown as solid black squares (experiment) and solid black circles (simulation).}
 \label{fgr:MA_properties_tuning}
\end{figure*}

\subsection{Internal energy alone does not dictate mechanical properties of amorphous solids:}

So far, through both bulk rheological measurements and microscopic non-affinity quantification, we have established that yielding in trained systems consistently occurs near the respective training amplitude. We now turn to explore the underlying energetic aspects of the system using simulations, since direct measurement of internal energy is not feasible in experiments.

It has been noted earlier that such oscillatory shear perturbation can anneal the system and take it to lower energy states. Therefore such protocols can also be considered as a annealing protocol facilitated by mechanical perturbation. Comparison reveals that such mechanical annealing (MA) are not as efficient~\cite{Yeh_PRL_2020,Bhaumik_PNAS_2021} as the thermal annealing (TA). The question we raise is, are these approaches (MA and TA) drive a poorly annealed system to similar regions in the energy landscape? This question has been schemtically depicted in Fig.~\ref{fgr:TA_MA_energy}(a). In Fig.~\ref{fgr:TA_MA_energy}(b) (right panel), we plot the potential energy of the system after the completion of mechanical annealing across a range of training amplitudes. With increasing training amplitude, the system undergoes mechanical annealing, which initially lowers the energy by guiding it into progressively more stable configurations. However, beyond a certain amplitude, the energy begins to rise again from the optimum. A nontrivial observation emerges here: although the energy changes non-monotonically with training amplitude ($\gamma_{\mathrm{T}}$), the yielding point remains sharply dependent on the training amplitude. This suggests that internal energy alone can not govern the yielding point of such trained material.

To further probe this idea, we also independently implemented a thermal annealing protocol designed to reduce the potential energy of the system without applying cyclic shear. Details of this procedure are provided in the Methods section. The left panel of Fig.~\ref{fgr:TA_MA_energy}(b) shows that thermal annealing indeed produces configurations with lower energy—consistent with the existing literature~\cite{Yeh_PRL_2020,Bhaumik_PNAS_2021}. However, as shown in Fig.~\ref{fgr:TA_MA_energy}(c), 
the stress versus strain and differential modulus curves for two mechanically annealed configurations and one thermally annealed configuration, all of which have comparable internal (potential) energy, have very different mechanical response during readout. For example, despite their similar energetic states, the three samples seem to have different shear modulus (G) and yield at different strain amplitudes (see also the inset of Fig.~\ref{fgr:TA_MA_energy}(c) for the shear modulus $G$). This clearly demonstrates that internal energy alone does not determine the yielding behavior, and mechanical training imprints additional structural memory that is not accessible through energy minimization via thermal annealing.

\subsection{Thermal and mechanical annealing control mechanical response  in very different ways}

In the previous section, we showed that systems prepared by thermal and mechanical annealing can have similar potential energy but can respond very differently to external deformation. We now explore how these two protocols independently influence different aspects of mechanical response of disordered solids and highlight their distinct roles in controlling shear response and their eventual mechanical failure aka yielding.

In Fig.~\ref{fgr:TA_MA_energy}(d), we compare the evolution of the yield strain under thermal and mechanical annealing protocols. The blue data points correspond to thermally annealed systems from a previous study~\cite{Misaki_PNAS_2018}, where the study explored a wide range of annealing states accessed  combining thermal treatment as well as SWAP algorithm~\cite{ninarello_PRX_2017}. In contrast, the red symbols represent our experimental and simulation data under mechanical annealing. We find that mechanical annealing enables tunability of the yield point over a wide range of perturbation amplitudes, whereas thermal annealing results in only modest shifts in the yield point, even under very high annealing conditions.

In Fig.~\ref{fgr:TA_MA_energy}(e), we present and compare the normalized linear shear modulus $\tilde{G}$ of the systems prepared by thermal annealing (from the Ref.~\cite{Misaki_PNAS_2018}) and mechanical annealing (from our experiments and MD simulations). The normalization is done with the most poorly annealed system in each case. Details of the linear modulus calculations are provided in the SI section S1. Interestingly, while mechanical annealing shifts the material's failure point to higher strains, it simultaneously leads to progressive softening as the training amplitude increases. This behavior is in contrast to thermal annealing, where the material show only a weak dependence of the yield strain on thermal annealing but becomes increasingly rigid with higher degree of annealing. These findings highlight that mechanical annealing enables access to a fundamentally different state of rigidity, offering opportunities for designing soft materials with tunable compliance and failure thresholds. This aspect will be discussed in some more details in the next section.

We further quantify the brittleness of the yielding transition by calculating the slope of the $G/G_0$ versus $\gamma$ curve beyond the yield point for each case. The detailed calculation is provided in the SI section S1. Since the absolute values of the post-yield slopes differ significantly between thermally and mechanically annealed systems, we normalize each slope by the respective maximum value attained in that annealing protocol, to allow for a direct comparison. The resulting normalized brittleness, denoted as $\tilde{m}$, is plotted in Fig.~\ref{fgr:TA_MA_energy}(f). For thermally annealed systems, brittleness increases monotonically with the degree of annealing. In contrast, mechanically annealed systems exhibit a non-monotonic trend, with maximum brittleness occurring at an intermediate training amplitude $\gamma_\mathrm{T} \sim 0.18$. Interestingly, this brittleness seems to be {\it{negetively correlated}} (see the SI section S6 for details) with the potential energy of the  mechanically annealed systems, the maximum brittleness coincides with the lowest internal energy. We also see a similar trend in the thermal annealing from our simulation (see the SI section S5), yet the data suggests energy alone can not determine the brittleness which reaffirm the ideas described in the previous section.

{\subsection{Role of hybrid annealing in tuning the mechanical properties}
}
From the previous section, we find that thermal and mechanical annealing modify the mechanical response of the system in fundamentally different ways. Thermal annealing only leads to a small increase in yield strain, while a pronounced enhancement of the linear shear modulus and brittleness. In contrast, mechanical annealing has very different effect: it significantly increases the yield point over a wide range, reduces the shear modulus with increasing training amplitude and brittleness changes in a non-monotonic way.

Because these two protocols tune different aspects of the mechanical response, we hypothesize that their combination can provide a powerful yet previously unexplored route to control the properties of the material into regimes that are inaccessible using either method alone.

To prove this hypothesis, we apply a thermo-mechanical annealing sequence followed by a uniform shear readout, as illustrated in Fig.~\ref {fgr:MA_properties_tuning}(a). The resulting responses, shown by filled green symbols in Fig.~\ref {fgr:MA_properties_tuning}(b), occupy a region of mechanical behavior that is inaccessible to either thermal (shaded blue) or mechanical (shaded red) annealing alone. This demonstrates that combining the two protocols enables access to previously unattainable combinations of shear modulus $G_0$ and yield strain $\gamma_{\mathrm{Y}}$. Note that this is just one of the possibilities of the combination of MA \& TA that we explored; optimization and control in such mixed protocol space is yet to be achieved.


To place our results in the broader context of yielding in disordered materials, we superimpose our measurements of the yield stress $\sigma_\mathrm{Y}$ and Young’s modulus $E$ onto the universal Ashby chart (reported in the Ref.~\cite{Cubuk_science_2017}) that contains most of the  naturally occurring or synthetic materials, as shown in Fig.~\ref {fgr:MA_properties_tuning}(c) and covers an enormous range, almost 15 orders of magnitude in Young's modulus. The procedures used to extract $E$ and measure $\sigma_\mathrm{Y}$ are detailed in the SI section S7. Consistent with the universal behavior of disordered solids, our untrained samples lie close to the established yield-strain line. Our data on thermal annealing also shows the same trend (see blue diamonds in Fig.~\ref {fgr:MA_properties_tuning}(c) and (d)). Strikingly, however, the effect of mechanical annealing is fundamentally different from the variations typically produced through changes in preparation or composition. As the training amplitude is increased, our data shift predominantly vertically in the $\sigma_\mathrm{Y}–E$ plane, as shown in Fig.~\ref {fgr:MA_properties_tuning}(d), rather than sliding along the universal line. This vertical evolution highlights a key outcome of our study: cyclic-shear enables precise and continuous control 
of mechanical response and opens up uncharted feature space in such trainable, disordered, mechanical metamaterials.
\\
\section{Discussions}
Our results demonstrate that mechanical annealing (MA), realized through cyclic shear, provides a powerful and unconventional means of tuning the yielding properties of amorphous solids. Unlike conventional thermal annealing (TA), which generally promotes increased rigidity and progressively solid-like behavior, MA enables a far more richer scenario--yield point changes across a broad range of strain amplitudes while simultaneously producing material softening and a non-monotonic change in brittleness. This counterintuitive combination underscores the fundamentally distinct nature of mechanical training protocols relative to thermal treatment in tuning material properties. To our knowledge, no previously reported conditioning or training protocol enables such broad and systematic control of the yield point in strain space. This yielding scenario serves almost as a mechanical analogue of a quote by  Friedrich Nietzsche ``Was mich nicht umbringt, macht mich st\"arker" or ``What does not kill me makes me stronger". Our results highlight the utility of MA as a powerful tool, not only to encode a robust memory but also to achieve customizable mechanical response in {\it{trainable amorphous matter}}.

In addition to contrasting MA and TA individually, our results suggest that combining the two protocols offers an even more powerful design strategy. While TA strengthens rigidity by increasing the elastic modulus, MA primarily enables targeted control of the yield point. We showed evidence that together, they can be used to produce materials that are both stronger (higher modulus) and more resistant to yielding (a significantly higher yield point), or alternatively, materials with tailored trade-offs between rigidity and flow-ability. Control over this duality highlights a new degree of freedom in the design principle of amorphous solids and soft meta-materials~\cite{Zaiser_NatRevPhy_2023}.

Such an approach also opens a broad research avenue: the possibility of engineering disordered systems through multi-step training protocols that couple thermal and mechanical routes in a strategic combination. This may allow programmable pathways~\cite{Falk_PNAS_2023} for tuning elasticity, plasticity, and flow across scales, with implications ranging from structural glasses and colloids to metallic alloys and biological materials. Overall our work reconfirms the idea that the yielding and rigidity of amorphous solids are not fixed intrinsic properties, but tunable outcomes of preparation history—offering a versatile framework for designing adaptive, multi-functional materials.

\section*{Acknowledgement}
P.A. acknowledges funding from the European Union’s Horizon Europe research and innovation programme under the Marie Skłodowska-Curie grant agree-
ment No 101149195.
P.A. also acknowledges the HPC facility in UGA, specifically the cluster ``DAHU" for computational purposes.  R.M. acknowledges support from the ANRF, India, through PMECRG (project ANRF/ECRG/2024/002036/PMS). SM acknowledges the Raman Research Institute for intramural funding.

\clearpage

\section{Methods}

\subsection {Experiments}

\subsubsection{Sample preparation}

We synthesize poly(N-isopropylacrylamide) (PNIPAM) particles using a one-pot free radical polymerization method\cite{brijitta}. After synthesis, microgel suspensions are purified through multiple cycles of centrifugation, decantation, and redispersion. The synthesized PNIPAM particles are characterized using dynamic light scattering (DLS) and scanning electron microscopy (SEM). After the final centrifugation step, the sediment is dried in an oven, crushed into a fine powder, and stored for subsequent sample preparation.

At 25$^{\circ}$C, the particles exhibit a mean hydrodynamic diameter $d_h = 0.63 \pm 0.057 \, \mu\text{m}$, with a volume phase transition temperature (VPTT) of approximately 33$^{\circ}$C.

For experimental use, dried PNIPAM particles are redispersed in Milli-Q water at various weight fractions ($wt\%$), defined as the ratio of the dried particles' weight to the total weight of the suspension. All the rheological experiments are performed at 25$^\circ$C.
A finite elastic modulus $G'$ in the linear regime (elastic modulus $G'$ is greater than the viscous modulus $G{''}$) is observed from 8 $wt\%$ onward, while training-induced yielding behavior becomes evident from 11 $wt\%$. 
The data presented in this study are based on a 20 $wt\%$ aqueous suspension of PNIPAM particles.

As the experiments are conducted at 25$^\circ$C, lower than the volume phase transition temperature (VPTT) of the PNIPAM particles, the particles remain in a swollen state\cite{Shibil_2023,Wu_2018,Lanzalaco_2023}. The density of the particles is close to that of \cite{Aangenendt_2017,Sbeih_2019}, making the suspension nearly optically transparent. To enhance light scattering for imaging purposes, 1 $wt\%$ polystyrene particles (diameter approximately 3 $\mu\text{m}$) are added to the suspension.

\subsubsection{Experimental setup}

Rheological measurements were performed using a stress-controlled MCR 702 rheometer (Anton Paar, Graz, Austria) equipped with sandblasted cone-plate geometries. The cone has a diameter of 25 mm, an angle of 2$^\circ$ and a truncation height of 106 $\mu\text{m}$. A Peltier-controlled bottom plate was used to maintain the temperature at 25$^\circ$C throughout the measurements. The sandblasted surfaces minimize wall slippage during shear experiments.

For in situ boundary imaging, the optical setup is integrated with the rheometer operated in a separate motor-transducer mode, where the shear is applied to the bottom plate and the cone remains fixed. During rheological measurements, the sample boundary is illuminated using an LED light source (Dolan-Jenner Industries), and scattered light is collected in the flow–gradient plane using a CCD camera (Lumenera) equipped with a 10× long working distance objective (Mitutoyo). The boundary images are processed to extract the flow field using Particle Image Velocimetry (PIV). PIV analysis is carried out in MATLAB using the mPIV toolbox developed by Nobuhito et al.(available at: \url{https://www.mathworks.com/matlabcentral/fileexchange/2411-mpiv}).

\subsubsection{Perturbation protocol}

Cyclic shear strain deformations are applied to train the system, followed by a uniform shear readout to probe its mechanical response. The training protocol follows a triangular waveform in strain, varying from $0 \rightarrow \gamma_{\mathrm{T}}\rightarrow 0$, with a constant shear rate magnitude of $0.01~\text{s}^{-1}$. This implies a forward strain rate of $+0.01~\text{s}^{-1}$ and a reverse rate of $-0.01~\text{s}^{-1}$ during each half-cycle. The training amplitude is varied from $\gamma_\mathrm{T}=0.02$ to $\gamma_\mathrm{T}=0.4$. In all cases, 50 training cycles are applied, which is sufficient for the system to reach a steady state prior to the readout phase.

\subsubsection{Non-affinity quantification}

To quantify non-affine displacements\cite{Maitri_2025}, we first measure the experimental velocity field $\mathbf{v}_E(x,y)$ from in situ boundary imaging using Particle Image Velocimetry (PIV). We focus on the
absolute value of the $x$ component of the velocity field $v^x_{E}(x, y)$, which lies along the direction of imposed shear.

To construct the affine flow field $v_A(y)$, we average the 
experimental flow field along the $x$ direction at the top $(y = d)$ and bottom $(y = 0)$:
\begin{equation*}
\langle v_E(d) \rangle_x = \frac{1}{L_x} \int_0^{L_x} v_E(x, d)\, dx
\end{equation*}
\begin{equation*}
\langle v_E(0) \rangle_x = \frac{1}{L_x} \int_0^{L_x} v_E(x, 0)\, dx
\end{equation*}
where $ L_x$ is the extent of the field of view along the $x$ direction (flow direction).

Using these two boundary values, we define the affine velocity field as a linear interpolation between the top and bottom planes:
\begin{equation*}
v_A(y) = \left( \frac{\langle v_E(d) \rangle_x - \langle v_E(0) \rangle_x}{d} \right) y + \langle v_E(0) \rangle_x
\end{equation*}
Note that $v_A(y)$ depends only on $y$, consistent with an ideal affine shear field.

Finally, the local non-affinity at each point $\Delta(x,y)$ is quantified as:
\begin{equation*}
\Delta(x, y) = \frac{\left| v_A(y) - v_E(x, y) \right|}{v_A(y)}
\end{equation*}
Here, $y$ is the velocity gradient direction, with shear applied from bottom to top. The field $\Delta(x,y)$ serves as a local measure of non-affine rearrangements within the material.

During the readout, images are captured at a frame rate of 1 frame per second (1 fps). Non-affine maps are generated by analyzing pairs of consecutive images. Since the shear rate during readout is $0.01~\text{s}^-1$, the non-affine field is computed over a strain increment of $\Delta\gamma=0.01$.

The mean non-affinity $\overline{\Delta}$ is calculated from each non-affine map as:
\begin{align*}
\overline{\Delta} = \overline{\Delta(x,y)} 
&=  \frac{1}{L_x L_y} \int_0^{L_x} \int_0^{L_y} \Delta(x, y)\, dy\, dx \\
&= \frac{1}{L_x L_y} \int_0^{L_x} \int_0^{L_y} 
\frac{\left| v_A(y) - v_E(x, y) \right|}{v_A(y)}\, dy\, dx
\end{align*}
where $L_x$ and $L_y$ are the dimensions of the field of view in the flow and velocity-gradient directions, respectively.

For all experiments, we also compute the initial mean non-affinity value, $\overline{\Delta}_0$, which represents the baseline non-affinity the system starts with during readout.  $\overline{\Delta}_0$ is obtained by averaging $\overline{\Delta}$ over the strain range up to approximately $\gamma=\gamma_\mathrm{T}/2$. The relative increase in non-affinity for each case is then quantified as:
\begin{equation*}
 \tilde{\Delta}=(\overline{\Delta}/\overline{\Delta}_0 - 1)
\end{equation*}

\subsection{Molecular Dynamics Simulations}
We perform two types of molecular dynamics (MD) simulations: athermal and thermal. Below, we describe each of them in detail.

\subsubsection{Athermal simulation}
In our athermal simulations, we consider a 50:50 bi-disperse mixture of spherical particles with diameters $\sigma$ and $1.4\sigma$. All quantities are expressed in reduced units, with $\sigma=1$ (the diameter of the smaller particle), $\epsilon=1$ (the interaction energy scale), and particle mass $m=1$, setting the units of length, energy, and mass, respectively. The corresponding time unit is $\tau=\sqrt{\sigma^2}/\epsilon$. Particles interact via a purely repulsive harmonic potential:
\begin{equation}
    V(r) =  K (r_0 - r)^2 H(r_0 - r),
\end{equation}
where $K=1$ is the stiffness constant which sets the energy scale $\epsilon$, $r$ is the center-to-center distance between two particles, and $r_0 = R_i + R_j$ is the sum of their radii. The Heaviside function $H(...)$ is used to indicate that the interaction is only active when particles overlap {\it{i.e.}} $r < r_0$. The system is enclosed in a three-dimensional cubic box with periodic boundary conditions applied in all directions.

We begin with an energy-minimized configuration at a target packing fraction $\phi = \phi_0$, and apply athermal quasistatic shear (AQS) by incrementally shearing the system in the $xy$ plane - corresponding to flow in the $x$-direction and gradient in the $y$-direction — in small strain steps $\Delta \gamma = 10^{-3}$, followed by energy minimization after each step. This process is continued up to a total shear strain $\gamma_T$.

Then, we reverse the strain direction, applying shear steps of $-\Delta \gamma$ and minimizing after each step until the system returns to the original simulation box. This complete back-and-forth shear cycle is repeated 200 times for each training amplitude $\gamma_T$.

Finally, we perform a readout by shearing the trained configuration again in the original direction, but now extending to much larger total strains. This allows us to observe the onset of yielding and determine the critical strain beyond which irreversible deformation occurs.

\subsubsection{Thermal Simulation}
In our thermal annealing simulations, we use the same bidisperse system as in the athermal case, but here we assign initial particle velocities drawn from a Gaussian distribution corresponding to a reduced temperature $T=0.1$. The system evolves under the velocity-Verlet algorithm in an NVT ensemble, with a Berendsen thermostat maintaining temperature control.
We gradually cool the system from $T=0.026$ down to $T=0.002$ in steps of $\Delta T=0.002$. Each temperature step is simulated for $200,000$ MD steps to ensure proper equilibration. After reaching the final temperature, we perform energy minimization using the conjugate gradient method to obtain a mechanically stable configuration. Here, temperature 
$T$ is expressed in reduced  units where $k_B=1$ and the energy scale $\epsilon$ is set by the interaction stiffness $K=1$. Therefore, $T$ corresponds to $k_B T/\epsilon$.

\bibliographystyle{apsrev4-2}
\bibliography{references}

\end{document}